
\magnification = \magstep1
\font\eightrm=cmr9
\long\def\fussnote#1#2{{\baselineskip=9pt
     \setbox\strutbox=\hbox{\vrule height 7pt depth 2pt width 0pt}%
     \eightrm
     \footnote{#1}{#2}}}
\baselineskip=18pt
\lineskip=2pt
\hsize=15 true cm
\vsize=22 true cm
\hoffset=0.75 true cm
\voffset=0.9 true cm
\vskip 1.5cm
\centerline{\bf The Shape of Inflated Vesicles}
\vskip 1cm
\centerline{
G. Gompper$^{(a)}$
and
D.M. Kroll$^{(b,}$\fussnote{$^{\dagger)}$}{Permanent address: Institut
f\"ur Festk\"orperforschung, KFA J\"ulich, Postfach 1913, 5170 J\"ulich,
Germany} }
\vskip 0.50cm
\centerline {$^{(a)}$ \sl Sektion Physik
der Ludwig-Maximilians-Universit\"at M\"unchen}
\centerline{\sl Theresienstr. 37, 8000 M\"unchen 2, Germany}
\vskip 0.50cm
\centerline {$^{(b)}$ \sl AHPCRC, University of Minnesota}
\centerline{\sl 1100 Washington Avenue South}
\centerline{\sl Minneapolis, MN 55415, USA}\vskip 1cm
\centerline{\bf Abstract}
\noindent
The conformation and scaling properties of self-avoiding fluid vesicles
with zero extrinsic bending rigidity subject to an internal pressure
increment $\Delta p>0$ are
studied using Monte Carlo methods and scaling arguments. With
increasing pressure, there is a first-order transition from a collapsed
branched polymer phase to an extended inflated phase. The scaling
behavior of the radius of gyration, the asphericities, and several
other quantities characterizing the average shape of a vesicle are
studied in detail. In the inflated phase, continuously variable
fractal shapes are found to be controlled by the scaling variable
$x=\Delta p N^{3\nu/2}$ (or equivalently, $y = {<V>}/ N^{3\nu/2}$),
where $N$ is the number of monomers in the vesicle and $V$ the
enclosed volume. The scaling behavior in the inflated phase is
described by a new exponent $\nu=0.787\pm 0.02$.

\vskip 1cm
\centerline{PACS numbers: 05.40.+j, 64.60.Fr, 87.22.Bt}
\vfill\eject
\vskip 1cm

\noindent
{\bf 1. Introduction}
\par In the absence of lateral tension, the shape of a fluid membrane
is determined by its elastic bending energy [1,2]. At length scales
small compared to the persistence length [3] $\xi_p = a_0
\exp({4\pi\over 3} \kappa/k_B T)$, where $a_0$ is a microscopic length
of the order of the monomer size and $\kappa$ is the bending rigidity,
equilibrium configurations minimize the elastic bending energy, and
thermal fluctuations can be ignored. At longer length
scales, however, fluctuations lead to a decrease  in the effective
bending rigidity [4] so that membranes of linear dimension $L\gg
\xi_p$ are expected to behave as crumpled objects characterized by
the absence of long-range orientational order. At these length scales,
where the effective rigidity is negligibly small, there is no energy
scale and fluctuations determine the equilibrium shape.

\par Closed membranes or vesicles, usually of spherical topology,
are of particular interest. For vesicles of radius $R \ll \xi_p$, it
has been shown [5-7] that the intricate shapes observed in experiment
correspond to equilibrium configurations which minimize the
Helfrich elastic bending energy [5]

$$ {\cal H} = 2\kappa \int dS \  (H-H_0)^2    \eqno(1) $$

\noindent subject to the constraints of constant area and enclosed
volume, where $H = {1\over 2} (1/R_1 + 1/R_2)$ is the mean curvature
(expressed in terms of the principle radii of curvature $R_1$, $R_2$),
and $H_0$ is the spontaneous curvature. Here, we are interested in the
limit of small bending rigidities [8], where $\xi_p \ll R$. This case has
been studied previously in considerable detail in ($d=2$)-spatial
dimensions, where the ``vesicle'' is a closed ring polymer [9-12]. The
shape of these polymer rings can be characterized by the thermal
averages of the eigenvalues $\lambda_2>\lambda_1$ of the moment of
inertia tensor. In particular, it has been shown [10] that the value of the
shape parameter $\Sigma \equiv <\lambda_1> / <\lambda_2>$ changes
continuously with the pressure increment
$$\Delta p = p_{in}-p_{out}.  \eqno(2) $$
For $\Delta p=0$, the polymer ring is a self-avoiding random walk,
with $\Sigma \simeq 0.39$. For $\Delta p \ne 0$, $\Sigma$ is a
universal function of the scaling variable
$$ x = p N^{2\nu}, \ \ \ {\rm with} \ \ \ p=\Delta p a^2/k_BT,
                                               \eqno(3) $$
where $N$ is the number of monomers in the ring, $a$ is a length scale
of the order of the interparticle spacing, and $\nu=3/4$ is the
self-avoiding random walk exponent in two dimensions which describes the
dependence of the radius of gyration, $R_g$, on the number of
monomers via
$$ <R_g^2>\ \sim N^{2\nu}.   \eqno(4)   $$
For $\Delta p>0$, the rings are inflated and approach a circular shape
for large $\Delta p$, so that $\Sigma(x)\to \Sigma^+ = 1$ for $x\to
\infty$.  For $\Delta p<0$, the rings collapse to a branched polymer
shape, with $\Sigma(x) \to \Sigma^- \simeq 0.23$ in the limit $x\to
-\infty$.

\par In this paper we consider the corresponding problem for
self-avoiding fluid vesicles in $d=3$ spatial dimensions. Our
analysis is based on an extensive Monte Carlo study of a simple
string-and-bead model for randomly triangulated vesicles. The surface
is modeled by a triangular network of $N$ hard-sphere particles of
diameter $a=1$ connected by tethers of maximum extension
$\ell_0=\sqrt{2}$ [13]. No explicit curvature term is included in the
model. The fluid character of the membrane is incorporated
by dynamically updating the triangulations. Details concerning
the simulation procedure can be found in Refs. 14-16.

\par Using this model we have recently shown [16-19] that the behavior
of fluid vesicles in $d=3$ spatial dimensions is drastically
different from that of ring polymers in two-dimensions. For example,
already at $\Delta p = 0$  a vesicle exhibits a branched polymer
conformation [16,17,20-23] characterized by the scaling laws [24]
$$ \eqalign{
     <R_g^2> &\sim N^{\nu_{bp}}, \cr
     <V> &\sim N^{\nu_{bp}},    \cr}  \eqno(5)  $$
with $\nu_{bp}=1$. As $\Delta p$ increases, there is a slow inflation
until at some critical pressure difference $\Delta p^*(N)>0$, there is
a {\it first-order} transition to a high pressure inflated phase
[18]. Increasing the pressure still further causes the vesicle to
slowly inflate until it becomes, ultimately, spherical. In this
paper, we characterize the inflated phase in some  detail. In Sec. 2,
scaling laws for the radius of gyration, the enclosed
volume, and the width of the vesicle wall are derived using a generalization
of de Gennes' blob picture. The scaling behavior of the critical
pressure difference $\Delta p^*(N)$ at the first-order transition is
discussed in Sec. 3, and data for the radius of
gyration and enclosed volume in the inflated phase are analyzed. A simple
argument is given which indicates that $\Delta p^* \sim N^{-1/2}$.
Sec. 4 contains a discussion of the behavior in the two-phase region,
and data taken in both the constant pressure and constant volume
ensembles are compared. A scaling analysis of various quantities
characterizing the vesicle shape in the inflated phase is presented
in Sec. 5; the paper closes with a brief summary and discussion.

\bigskip
\noindent
{\bf 2. Scaling}
\smallskip
The scaling behavior in the inflated phase can be derived using a
simple generalization of the blob picture of de Gennes [11,25-27,18].
Consider a  $(d-1)$-dimensional crumpled membrane embedded in
$d$-dimensional space.  In the blob picture, one envisages the
membrane breaking up into a set of $N_b$ weakly stretched blobs of
area proportional to $\xi_\sigma^{d-1} = k_BT/\sigma$, where
$\xi_\sigma$ is the tensile length, under the  influence of a uniform
extensional tension $\sigma$ applied to the  membrane perimeter. Since
the blobs themselves are only weakly  stretched, it is further assumed
that
$$ \xi_\sigma^{d-1} \sim M_b^\nu,   \eqno(6) $$
where $M_b=N/N_b$ is the number of monomers in a blob. It is then
argued that the blobs become independent on length scales much larger
than $\xi_\sigma$ so that the total projected surface area $<A_b>$ is
proportional to $\xi_\sigma^{d-1}(N/M_b)\sim N\sigma^{1/\nu-1}$.  If
an inflated vesicle is regarded as a (hyper-) spherical bubble of
radius $R$ with a surface tension $\gamma$, one has $\Delta p =
2\gamma/R$. Identifying the surface tension with the stretching
tension $\sigma$,  and taking $<A_b>\sim R^{d-1}$, we arrive at the
scaling form
$$ R_g \sim p^{\omega} N^{\nu_+}  \eqno(7) $$
for the radius of gyration of an inflated vesicle, with $p=\Delta p
a^d/k_BT$, and the exponents
$$ \eqalign{
    \omega &= {1-\nu \over d\nu-1} \cr
    \nu_+  &= {\nu \over d\nu-1}. \cr } \eqno(8) $$
$\nu$ is a new exponent which characterizes
the scaling behavior within a blob in the inflated phase.

\par Eq. (7) is the large extension limit of a more general scaling
form. Assuming that the radius of gyration $R_g$ and the tensile
strength  $\xi_\sigma$ are the only two relevant length scales in the
problem, it is natural to expect the projected area $A_b$ to scale as
$$ A_b =  R_g^{d-1}\Omega_0(\xi_\sigma/R_g) =
                         N^\nu \Omega(\sigma N^\nu),  \eqno(9)  $$
where the last relation follows from the definition $R_g^{d-1}\sim
N^\nu$, and $\xi_\sigma^{d-1}\sim \sigma^{-1}$.
For large $\sigma$, $A_b$ should vary {\it linearly} with $N$, so
that
$$ \Omega(z) \sim z^{{1\over \nu}-1} \ \ \ \ {\rm for}
                 \ \ \ \ z\to \infty.     \eqno(10) $$
This implies $A_b \sim N \sigma^{(1-\nu)/\nu} $ in this limit.
For an inflated vesicle we have (see discussion preceding Eq.(7))
$ \sigma \sim p R_g  $ so that
$$ A_b \sim R_g^{d-1} \sim N (p R_g)^{(1-\nu)/\nu}    \eqno(11)  $$
in the large extension limit.
Solving this equation for $R_g$, we arrive at (7) with the
exponents (8).  From (9), we see
that the full scaling form in the inflated phase is
$$ <R_g^2> = p^{2\omega} N^{2\nu_+} \Xi_p( p N^{d\nu/(d-1)} ).
                                                   \eqno(12) $$
Similarly, the mean volume can be shown to scale as
$$ <V> = p^{d\omega} N^{d\nu_+} \Gamma_p( p N^{d\nu/(d-1)} ).
                                                   \eqno(13) $$
Finally, Eq. (13) suggests that the scaling behavior of the mean
square radius of
gyration of vesicles fluctuating at fixed enclosed {\it volume}
(instead of at fixed pressure) scales as
$$ <R_g^2> = V^{2/d} \Xi_V( V N^{-d\nu/(d-1)} ).  \eqno(14) $$
in the single phase inflated regime.
Note that for $d=2$, these results are identical with the those of
Ref. 11 for polymer rings.

\par We can go further and calculate the radial extension of the
highly convoluted surface of the vesicle. The radial thickness $\xi$
should scale like the projected area, Eq.(9),
$$ \xi^{d-1} = N^\nu Q( \sigma N^\nu ) .  \eqno(15)  $$
In the large tension limit, $\xi$ should become {\it independent}
of $N$, so that
$$ Q(z) \sim z^{-1} \ \ \ \ \ {\rm for} \ \ \ z\to\infty.
                                                   \eqno(16) $$
This implies that
$$ \xi \sim (pR_g)^{-1/(d-1)} \sim p^{-\nu_+}N^{-\nu_+/(d-1)}
                                                    \eqno(17) $$
in this limit.
The thickness $\xi$ described by (17) is just the radius (6) of the
blobs in the blob picture. It should therefore be considered to be the
intrinsic thickness of the vesicle walls. In particular, it does not
take into account  the effect of {\it capillary waves} on length
scales larger than $\xi$. These excitations can be easily estimated
using an effective Hamiltonian for a {\it nearly flat} interface of
projected area $A_b$ (with periodic boundary conditions). We introduce
a coarse grained variable $u({\bf r})$ which measures the distance of
the membrane from some reference plane. On length scales larger than
$\xi$, the lateral tensions dominates the fluctuations, so that
$$ {\cal H}_{eff} = {\sigma\over 2} \int_{A_b} d^{d-1}r (\nabla u)^2,
                                              \eqno(18) $$
with an implicit short length scale cutoff $\xi$. The amplitude of the
fluctuations is calculated with the help of the equipartition theorem,
as usual, with the result
$$ \eqalign{
 <u^2> &= \int_{R_g^{-1}<q<\xi^{-1}} {d^{d-1} q \over (2\pi)^{d-1} }
                                 {k_B T\over \sigma q^2}  \cr
       &\sim \cases{ {k_B T\over \sigma} R_g^{3-d}
                                         &for $d<3$ \cr
          {k_B T\over \sigma} \ln(R_g/\xi) &for $d=3$ \cr} \cr }
                                            \eqno(19)  $$
Inserting the results for $\sigma$ and $R_g$, we arrive at
$$  \sqrt{<u^2>\over <R_g^2>} \sim \cases{
      \left(p N^{d\nu/(d-1)} \right)^{-1/(d\nu-1)}
                                  &for $d<3$ \cr
           \left(p N^{3\nu/2} \right)^{-1/(3\nu-1)}
                    \sqrt{ \ln\left( p N^{3\nu/2} \right) }
                                  &for $d=3$ \cr}
                                                 \eqno(20)  $$
so that capillary wave broadening dominates the thickness of the
vesicle walls for all $d\le 3$. However, for $d=3$, which is the case
of interest, the effect of capillary waves is very weak, and
can only be observed for very large vesicles.

\bigskip
\noindent
{\bf 3. Radius of Gyration and Volume}
\smallskip
The behavior of the average volume $<V>$ of a vesicle as a function
of $\Delta p$ is shown in Fig. 1 for three different system sizes,
$N=47$, 127, and 247. Each data point corresponds to an
equilibrium average over $20-40$ million Monte Carlo steps per
monomer (MCS). The evolution of a discontinuity in $<V>$ with
increasing $N$ is clearly visible. Since
the transition pressure is difficult to determine from $<V(p)>$
itself, we have also calculated the probability distribution $P(V)$.
For $N=47$, the distribution has a single sharp peak for low and high
pressures, but broadens and exhibits a weakly bimodal structure near
$\Delta p^*=1.05\pm 0.05$, which we
identify with the location of the transition. The distribution for
$N=127$ is shown in Fig. 2 for various pressures near the transition.
For $\Delta p=0.50$, a bimodal distribution is found, with a somewhat
larger peak at large volumes. We estimate that the transition occurs
at $\Delta p^* = 0.47\pm 0.02$ in this case. For $N=247$ it is
more difficult to determine the location of the transition, since the
vesicle jumps back and forth between the inflated and the branched
polymer states so rarely that the equilibrium probability distribution
for $V$ could not be determined with sufficient accuracy. However, by
studying the time dependence of $V$ for $\Delta p=0.30$ and $\Delta p
= 0.35$, see Fig. 3, we conclude that the transition occurs at $\Delta
p^* = 0.34\pm 0.01$ for $N=247$. From the three vesicles sizes we have
studied, it is difficult to determine the $N$-dependence of $\Delta
p^*$.  Nevertheless, we present a simple argument at the end of this
Section which suggests that $\Delta p^*\sim N^{-1/2}$. This is in
excellent  agreement with our estimates for the transition pressures
quoted above  for $N=127$ and $N=247$.

\par As already shown in Ref. 18, the data for $<V>$ in the inflated
phase ($\Delta p>\Delta p^*$) scale according to (13) with an
exponent $\nu=0.787\pm0.020$. In fact, all our data for both the
volume and the mean square radius of gyration in the {\it inflated}
regime scale for this value of $\nu$. The scaling function $F$ for the
volume,

$$ F( p N^{3\nu/2} ) = <V>/(p^{3\omega} N^{3\nu_+}),  \eqno(21) $$

and the scaling function $\Xi_V( <V>/ N^{3\nu/2} )$ for the mean
square radius of gyration, see Eq. (14), are shown in Fig. 4. At very
large inflations, scaling breaks down. The deviation of the last
(large $\Delta p$) data point for each vesicle size in Fig. 4a is due
to this effect and thus delimits the scaling region.  For large
scaling argument, $F(y)$ must approach a constant value. It is
interesting to note that this occurs rather rapidly,  so that
significant deviations can be observed only for small pressures and
vesicle sizes. In fact, for $N\to\infty$, this asymptotic limit is  is
attained in all of the inflated phase, since as argued below,
$\Delta p^*$ scales as $N^{-1/2}$ for large $N$ so that
$\Delta p N^{3\nu/2}\to\infty$ for all $\Delta p>\Delta p^*$.

\par In Fig. 4b, as well as in Fig. 7 below, we have replaced $N$ by
$(N-N_0)$ in the scaling function in order to hasten convergence when
$N\to\infty$. This parameterization incorporates the leading
corrections to scaling.

\par Given these results, there is a simple argument which indicates
that the transition between the collapsed and inflated phases
occurs at a critical pressure difference $\Delta p^*$ which scales as
$$\Delta p^*\sim N^{-1/2}, \eqno(22)$$
{\it independent of the value of $\nu$}. For $\Delta p=0$, the vesicle
exhibits a branched polymer conformation. To leading order, the
configurational entropy $S(N)$
of a branched polymer of length $N$ is given $S(N) = N\ln(z)-\theta \ln(N)$,
where $z$ is a nonuniversal quantity independent of $N$ and
$\theta=3/2$ [24]. The free energy in the deflated phase therefore
scales as
$$\beta{\cal F}_- = - N \ln(z) - c\Delta p N \eqno(23) $$
to leading order in $N$ and $\Delta p$, where $c$ is a constant of order
unity. In writing (23)
we have included the leading pressure dependent contribution [18]
$\Delta p <V>\sim \Delta p N$. In the inflated phase, on the other hand,
we know that $<V>=-\partial{\beta{\cal F}}/\partial\Delta p$ scales
as $V_0^+\Delta p^{3\omega}N^{3\nu_+}$ [28]. The leading contribution to the
free energy in the inflated phase is, therefore,
$$\beta{\cal F}_+ = - {V_0^+ \over{3\omega+1}}
                      (\Delta p)^{3\omega+1}N^{3\nu_+}.  \eqno(24)$$
The transition occurs when the free energies of the deflated and
inflated phases become equal, i.e. for
$$N\sim (\Delta p^*)^{3\omega+1}N^{3\nu_+}, \eqno(25)$$
or, equivalently, $\Delta p^*\sim N^{-1/2}$, independent of $\nu$.

\bigskip
\noindent
{\bf 4. Two-Phase Region}
\smallskip
We have performed simulations in both the constant-pressure and
\break\hfill constant-volume ensembles. Our motivation is twofold.
First, we want to check the equivalence of the two ensembles for large
vesicles. This is important because experiments are usually performed
at constant volume [7,29]. Second, we want to see what a vesicle looks
like in the `two-phase' region. Namely, is there two-phase coexistence
between a branched polymer region and an inflated part of the vesicle?

\par A few typical configurations for $N=247$ and $V=120$ are shown
in Fig. 5. They demonstrate that the vesicle does indeed exhibit
two-phase coexistence, with a tree-like part attached to a
quasi-spherical region. Also, we find that $N=247$ is large enough to
give essentially identical results in the inflated phase for the two
ensembles. This can be seen in Fig. 4b, where the scaling function of
the mean square radius of gyration is plotted for a range of volumes
covering the
inflated phase, the two-phase region, as well as the
collapsed, branched polymer phase. Data obtained in both ensembles are
included  in this figure. Note that the data appear to scale over the
whole range  of volumes. We will see this remarkable behavior for
other scaling functions in Section 5. We want to emphasize, however,
that no true scaling can occur in the two-phase region.

\par For fully extended spherical vesicles, the volume and the radius
of gyration are related by $V={4\over 3}\pi R_g^3$. Thus,
$$ \tilde \pi \equiv {3\over 4} <V> <R_g^2>^{-3/2}      \eqno(26)  $$
is a measure for the deviation of the shape of the vesicle from an
ideal sphere. Our results for $\tilde\pi( <V> N^{-3\nu/2} )$ are shown in
Fig. 6. For $V$ large at fixed $N$, $\tilde\pi\to \pi$, as expected.
Fig.6 shows again that the data appear to scale not only in the
inflated phase, but in the two-phase region as well.

\bigskip
\noindent
{\bf 5. Mean Vesicle Shapes in Inflated Phase}
\smallskip
In order to characterize the shape of the vesicles, we have studied
the eigenvalues $\lambda_1<\lambda_2<\lambda_3$ of the moment of
inertia tensor,
$$ {\cal T}_{\alpha,\beta} =
    {1\over 2 N^2} \sum_{i,j} {q_i\over 6} {q_j\over 6}
    [r_i^\alpha-r_j^\alpha] [r_i^\beta-r_j^\beta],   \eqno(27) $$
where ${\bf r}_i$ is the position of monomer $i$, and $q_i$ is its
coordination number. In particular, we have analysed the
asphericities, $\Gamma_1=<\lambda_1/\lambda_3>$,
$\Gamma_2=<\lambda_2/\lambda_3>$, as well as the quantities [17,30,31]
$$\Delta_d = {d\over{d-1}} {<{\rm Tr}\ \hat{\cal T}^2>\over<({\rm Tr}
         \ {\cal T})^2>}             \eqno(28)$$
and
$$S_d = {d^2\over{(d-1)(d-2)}}{<{\rm Tr}\ \hat{\cal T}^3>\over
            <({\rm Tr} \ {\cal T})^3>},   \eqno(29) $$
where $\hat{\cal T}_{\alpha\beta} = {\cal T}_{\alpha\beta} - \bar\lambda
\delta_{\alpha\beta}$, and $\bar\lambda = {1\over d}\sum_{i=1}^d
\lambda_i$. $0\le\Delta_d\le1$ is a normalized measure of the anisotropy.
The sign of $-{1\over{(d-1)^3}}\le S_d\le 1$ determines whether the
object is oblate ($S_d<0$), or prolate ($S_d>0$); its magnitude is a
measure of the strength of the anisotropy. For $d=3$, one has
$$ \Delta_3 = {< \lambda_1^2 + \lambda_2^2 + \lambda_3^2 -
(\lambda_1\lambda_2 + \lambda_2\lambda_3 + \lambda_3\lambda_1) > \over
 < (\lambda_1 + \lambda_2 + \lambda_3)^2>}   \eqno(30)  $$
and
$$ S_3 = {<(\lambda_1-\bar\lambda)(\lambda_2-\bar\lambda)
          (\lambda_3-\bar\lambda)>\over 2<\bar\lambda^3>}. \eqno(31) $$

\par Let us first consider the averaged eigenvalues $<\lambda_i>$,
$i=1,2,3$, of the moment of intertia tensor.
In the inflated phase, the deviation of the vesicle from a completely
spherical shape is determined by the fluctuations studied in Section
2. To leading order, we therefore expect
$$ <\lambda_i> \simeq {1\over3} \left[ \sqrt{<R_g^2>}
         + {1\over 2}c_i \sqrt{<u^2>} \right]^2.       \eqno(32) $$
so that [32]
$$ {3<\lambda_i> \over <\lambda_1+\lambda_2+\lambda_3>} (y) =
   1 + c_i y^{-1/(3-3\nu)},
                                                     \eqno(33) $$
with $y=<V> N^{-3\nu/2}$.
Since $\sum_i<\lambda_i>=<R_g^2>$, one has $\sum_i c_i = 0$.
With the critical exponent $\nu=0.787$ determined above, we therefore
expect $[3<\lambda_i>/<R_g^2>-1]$ to decay asymptotically as
$y^{-1.565}$. The Monte Carlo data are shown in Fig. 7. The
eigenvalues $<\lambda_1>$ and $<\lambda_3>$ exhibit the expected
asymptotic power law behavior with an exponent which is very close to
the value quoted above. $[1-3<\lambda_2>/<R_g^2>]$, on the other hand,
decays much faster for large $y$, approximately as $y^{-3.05}$.
Indeed, the data are consistent with $c_2\simeq 0$. Since we expect,
in general, that corrections to spherical shape in the large inflation
limit can be expressed as a power series in $\sqrt{<u^2>/<R_g^2>}$, the
asymptotic behavior in this case should be given by the next to leading
term, which is $y^{-2/(3-3\nu)}=y^{-3.13}$, in good agreement with
our numerical result. From Fig. 7, the constants $c_i$ in Eq. (33) are
found to be
$$ c_1 = -0.012, \ \ \ c_2 = 0, \ \ \ {\rm and} \ \ \ c_3 = 0.013,
                                                     \eqno(34) $$
consistent with the requirement $\sum_ic_i=0$.
\par The asphericities can be analysed in a similar way. Under the
assumption that $\Gamma_{1,2} = <\lambda_{1,2}/\lambda_3> \simeq
<\lambda_{1,2}>/<\lambda_3>=\Sigma_{1,2}$, we obtain [32]
$$ 1-\Gamma_{1,2}(y) = - (c_{1,2}-c_3) y^{-1/(3-3\nu)}
                                                       \eqno(35) $$
with $y=<V> N^{-3\nu/2}$. It has been shown in Ref. 17 that for
$\Delta p=0$  the $\Gamma_i$ and $\Sigma_i$ differ by very little, so
that this assumption seems justified. The data shown in Fig. 8 are in
reasonable agreement with the scaling form (35), with an amplitude
which is indeed close to the difference of the $c_i$'s. Note, however,
that Eq. (35) is only an approximation, and that in general, one would
expect $1-\Gamma_i$ to scale with an independent exponent. In fact, it
appears that the asymptotic slope of $1-\Gamma_1$ does differ
somewhat from (35).

\par Finally, the shape parameters $\Delta_3$ and $S_3$ are shown in
Fig. 9. The data again fall onto a single curve when the scaling
variable $y=<V> N^{-3\nu/2}$ is used. The asymptotic behavior in this
case is found to be
$$ \eqalign{
\Delta_3(y) &\sim y^{-\gamma_\Delta}   \cr
 S_3(y)     &\sim y^{-\gamma_S},        \cr }   \eqno(36)  $$
with $\gamma_\Delta\simeq 3.31$, and $\gamma_S\simeq 6.22$.
Since these quantities are also averages of the eigenvalues
$\lambda_i$, it is tempting to conjecture that
$$ \gamma_\Delta \simeq 2/(3-3\nu) \ \ \ {\rm and} \ \ \
   \gamma_S \simeq 4/(3-3\nu).                         \eqno(37) $$
Note that $S_3>0$, but very small, so that the vesicles are prolate on
average.

\bigskip
\noindent
{\bf 6. Summary and Discussion}
\smallskip
The scaling behavior of fluid vesicles subject to an internal pressure
increment $\Delta p>0$ has been studied. Clear evidence has been
presented for a
phase transition from a low-pressure branched-polymer phase to a
high-pressure inflated phase. Several quantities characterizing the
vesicle shape, such as the average eigenvalues of the moment of
intertia tensor and the asphericities, have been determined
using Monte Carlo methods. In the inflated phase, all quantities
are found to be universal functions of the scaling variable
$x=\Delta p N^{3\nu/2}$ (or equivalently, $y = {<V>} N^{-3\nu/2}$),
with $\nu=0.787\pm 0.02$. The asymptotic scaling behavior for large
inflation can be understood using simple scaling arguments.

\par Much still needs to be learned regarding the properties of the
inflated phase, however. Although we feel that the nature of the
transition between the branched polymer phase and the inflated phase
is well  established, the relatively modest system sizes which can
currently be  studied make it difficult to determine $\nu$ precisely,
or even exclude the possibility that the behavior we observe in the
inflated phase is dominated by crossover effects. Nevertheless, the
fact that   arguments based on the blob picture of de Gennes yield
such a concise  and consistent description of the observed scaling
behavior makes us believe that the scenario described above is
basically correct.  In any case, a better theoretical understanding of
the nature of the inflated phase is necessary if we are to appreciate
the implications of these results.

\bigskip
\noindent {\bf Acknowledgement}: This work was supported in part by
Army Research Office contract number DAAL03-89-C-0038 with the
Army High Performance Computing Research Center at the University of
Minnesota, NATO grant CRG910156, and the Deutsche Forschungsgemeinschaft
through Sonderforschungsbereich 266.
\vfill\eject

\centerline{REFERENCES}
\smallskip
\item{[1]} {\it Statistical Mechanics of Membranes and Surfaces},
edited by D.R. Nelson, T. Piran, and S. Weinberg (World Scientific, 1989).
\smallskip
\item{[2]} R. Lipowsky, Nature {\bf 349}, 475 (1991).
\smallskip
\item{[3]} P.G. de Gennes and C. Taupin, J. Chem. Phys. {\bf 86}, 2294
(1982).
\smallskip
\item{[4]} L. Peliti and S. Leibler, Phys. Rev. Lett. {\bf 54},
1690 (1985).
\smallskip
\item{[5]} W. Helfrich, Z. Naturforsch. {\bf 28c}, 693 (1973);
H.J. Deuling and W. Helfrich, J. Phys. (Paris) {\bf 37}, 1335 (1976).
\smallskip
\item{[6]} S. Svetina and B. Zeks, Eur. Biophys. J. {\bf 17}, 101
(1989).
\smallskip
\item{[7]} K. Berndl, J. K\"as, R. Lipowsky, E. Sackmann, and U.
Seifert, Europhys. Lett. {\bf 13}, 659 (1990).
\smallskip
\item{[8]} Synthetic lipid bilayer vesicles containing a small
percentage of bola lipid are known to have bending rigidities of the
order of $k_B T$, or less, see
H.P. Duwe, J. K\"as, E. Sackmann, J. Phys. (Paris)
{\bf 51}, 945 (1990).
\smallskip
\item{[9]} S. Leibler, R.R.P. Singh, and M.E. Fisher, Phys. Rev. Lett.
{\bf 59}, 1989 (1987).
\smallskip
\item{[10]} C.J. Camacho and M.E. Fisher, Phys. Rev. Lett. {\bf 65}, 9
(1990).
\smallskip
\item{[11]} A.C. Maggs, S. Leibler, M.E. Fisher, and C.J. Camacho,
Phys. Rev. A{\bf 42}, 691 (1990).
\smallskip
\item{[12]} C.J. Camacho, M.E. Fisher, and R.R.P. Singh, J. Chem.
Phys. {\bf 94}, 5693 (1991).
\smallskip
\item{[13]} This value of $\ell_0$ is small enough to ensure self-avoidance
while still yielding bond-flip acceptance rates ($\sim 8\%$
for $\Delta p=0$ and $\sim 2-3\%$ for large $\Delta p$) which are large
enough to allow the monomers to diffuse in the membrane surface.
\smallskip
\item{[14]} A. Billoire and F. David, Nucl. Phys. B{\bf 275}, 617
(1986); D.V. Boulatov, V.A. Kazakov, I.K. Kostov, and A.A. Migdal,
Nucl. Phys. B{\bf 275}, 641 (1986).
\smallskip
\item{[15]} J.-S. Ho und A. Baumg\"artner, Europhys. Lett. {\bf 12},
295 (1990); A. Baum-
\break\hfill\noindent
g\"artner and J.-S. Ho, Phys. Rev. A{\bf 41}, 5747 (1990).
\smallskip
\item{[16]} D.M. Kroll and G. Gompper, Science {\bf 255}, 968 (1992).
\smallskip
\item{[17]} D.M. Kroll and G. Gompper, Phys. Rev. A (to appear).
\smallskip
\item{[18]} G. Gompper and D.M. Kroll, Europhys. Lett. (to appear).
\smallskip
\item{[19]} D.M. Kroll and G. Gompper, in {\it Statistical
Thermodynamics and Differential Geometry of Microstructured
Materials}, edited by H.T. Davis and J.C.C. Nitsche (Springer,
New York, 1992).
\smallskip
\item{[20]} U. Glaus, Phys. Rev. Lett. {\bf 56}, 1996 (1986);
J. Stat. Phys. {\bf 50}, 1141 (1988).
\smallskip
\item{[21]} J. O'Connell, F. Sullivan, D. Libes, E. Orlandini,
M.C. Tesi, A.L. Stella, and T.L. Einstein, J. Phys. A: Math. Gen.
{\bf 24}, 4619 (1991).
\smallskip
\item{[22]} D. Boal and M. Rao, Phys. Rev. A{\bf 45}, R6947 (1992).
\smallskip
\item{[23]} C.F. Baillie and D.A. Johnston, Phys. Lett. B{\bf 283}, 55
(1992).
\smallskip
\item{[24]} G. Parisi and N. Sourlas, Phys. Rev. Lett. {\bf 46},
871 (1981).
\smallskip
\item{[25]} P.-G. de Gennes, {\it Scaling Concepts in Polymer
Physics} (Cornell University Press, Ithaca, New York, 1979).
\smallskip
\item{[26]} P. Pincus, Macromolecules {\bf 9}, 386 (1976).
\smallskip
\item{[27]} R. Lipowsky and A. Baumg\"artner, Phys. Rev. A{\bf 40},
2078 (1989).
871 (1981).
\smallskip
\item{[28]} As can be seen from Fig. 4a, this is a good approximation only
for the two larger system sizes we studied. For $N=47$ there appear to be
finite $\Delta p N^{3\nu/2}$ corrections (see Eq. (13)). We will neglect
this effect in the current discussion.
\smallskip
\item{[29]} E. Evans and W. Rawicz, Phys. Rev. Lett. {\bf 64}, 2094
(1990).
\smallskip
\item{[30]} J.A. Aronovitz and D.R. Nelson, J. Phys. France {\bf 47},
1445 (1986).
\smallskip
\item{[31]} J.A. Aronovitz and M.J. Stephen, J. Phys. A: Math. Gen.
{\bf 20}, 2539 (1987).
\smallskip
\item{[32]} In Eq. (33), logarithmic corrections of the form $\sqrt{\ln(y)}$,
have been ignored, see Eq. (19). For the limited range of values of
the scaling variable $y$ accessible in our simulations, such
corrections cannot be distinguished from simple power laws.

\vfill\eject
\centerline{FIGURE CAPTIONS}
\smallskip
\item{1.)} The average volume $<V>$ as a function of the pressure
increment $p$, for $N=47$ ($\diamond$), $N=127$ ($\times$), and
$N=247$ ($\bar\sqcup$). $p=\Delta p a^3/k_BT$, where the constant $a$ is
of the order of the interparticle spacing; for convenience we set
$a^3/k_BT=1$ here and in the following.
\smallskip
\item{2.)} Probability distribution $P(V)$ for $N=127$, at $p=0.40$
(dashed line), $p=0.50$ (full line), and $p=0.60$ (dotted line).
\smallskip
\item{3.)} Volume $V$ vs. number of Monte Carlo steps per monomer, for
$N=247$, $p=0.30$ and $p=0.35$. In the case $p=0.30$, the data are
shifted by $\Delta V=-40$ in order to separate the two curves.
\smallskip
\item{4.)} Scaling functions for (a) the volume, $F=<V> p^{-3\omega}
N^{-3\nu_+}$, as a function of the scaled pressure $p N^{3\nu/2}$, and
(b) the mean square radius of gyration, $\Xi_V = <R_g^2><V>^{-2/3}$,
as a function
of the scaled volume $<V> (N-N_0)^{-3\nu/2}$, with $N_0=13$. Data
for both the constant-pressure ensemble with $N=47$ ($\diamond$),
$N=127$ ($\times$), and $N=247$ ($\bar\sqcup$), as well as the constant
volume ensemble with $N=247$ ($\circ$) are plotted.
\smallskip
\item{5.)} Configurations of a vesicle in the two-phase region with
$N=247$ monomers and volume $V=120$, after $4, 8, 12, 16$ and $20$
million MCS.
\smallskip
\item{6.)} The volume to area ratio $\tilde\pi( <V> N^{-3\nu/2} )$ of an
inflated vesicle. Data for both the constant-pressure ensemble
with $N=47$ ($\diamond$), $N=127$ ($\times$), and $N=247$
($\bar\sqcup$), as well as the constant volume ensemble with $N=247$
($\circ$) are plotted.
\smallskip
\item{7.)} Scaling plot of the eigenvalue ratios
$3<\lambda_i>/<\lambda_1+\lambda_2+\lambda_3>$ of the moment of
inertia tensor as a function of $<V> (N-N_0)^{-3\nu/2}$, with
$N_0=5$. The solid lines in (a) and (c) indicate the asymptotic scaling
behavior (33) with the amplitudes $c_3=0.013$ and $c_1=-0.012$,
respectively. Data for both the constant-pressure ensemble with
$N=47$ ($\diamond$), $N=127$ ($\times$), and $N=247$ ($\bar\sqcup$),
as well as the constant volume ensemble with $N=247$ ($\circ$) are
plotted.
\smallskip
\item{8.)} Scaling plot of the asphericities $\Gamma_1$ and
$\Gamma_2$ as a function of $<V> N^{-3\nu/2}$. The solid lines indicate
the scaling behavior (35) with the amplitudes $(c_3-c_1)=0.025$ and
$(c_3-c_2) = c_3 = 0.013$. Data for both the constant-pressure
ensemble with $N=47$ ($\diamond$), $N=127$ ($\times$), and $N=247$
($\bar\sqcup$), as well as the constant volume ensemble with $N=247$
($\circ$) are plotted.
\smallskip
\item{9.)} Scaling plot of $\Delta_3$ and $S_3$ as a function of
$<V> N^{-3\nu/2}$. Data for both the constant-pressure ensemble with
$N=47$ ($\diamond$), $N=127$ ($\times$), and $N=247$ ($\bar\sqcup$),
as well as the constant volume ensemble with $N=247$ ($\circ$) are
plotted.

\end